# Embodying Newtonian Mechanics

*Carolyn Sealfon, Minerva University*
*csealfon@metalearning.ca*

Wellness and mindfulness act as buzzwords these days, often seen as separate from physics. Yet we know they are important, and everything is related to physics! In this article, we will consider a few simple classroom activities that can both help students internalize the basic physics of forces and motion and also help facilitate well-being in our classes.

As Brookes, Etkina, and Planinšič argue, it is helpful to clarify our intentions as physics instructors and to align our instruction with our intentions. One valuable intentionality they propose is, "The way in which students learn physics should enhance their well being." This intentionality underlies the renowned Investigative Science Learning Environment (ISLE) approach to teaching high-school and first-year university physics (Brookes et al., 2020).

As physics instructors, one way we can integrate well-being and physics is through inviting students to feel the physics that they are learning in their bodies. When we are asked, "What do you teach?" we usually respond, "physics," when really the answer is "human beings". Very often, especially in physics, we and our students can get "stuck in our heads" and almost forget we are whole human beings.

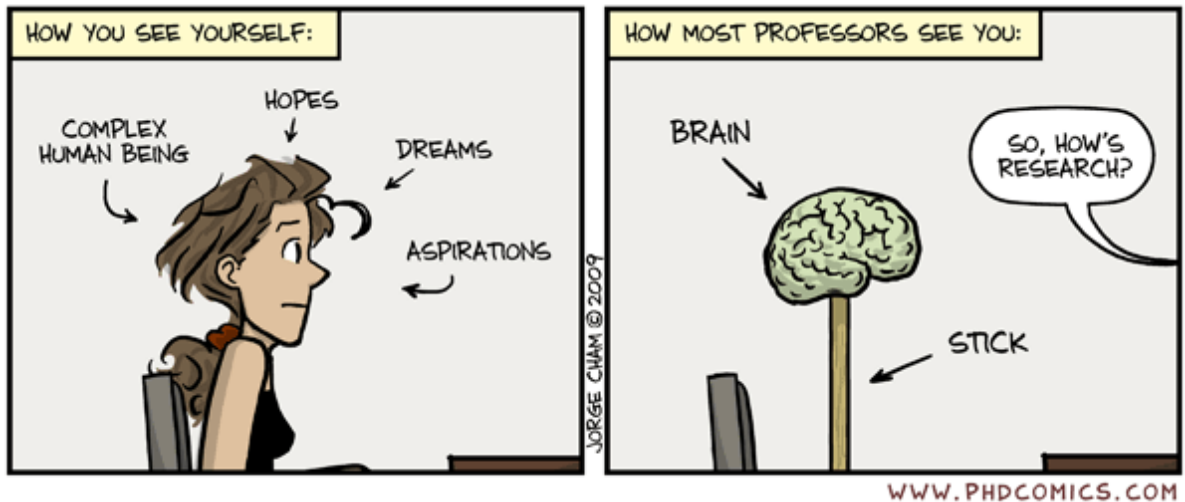

Image from "Piled Higher and Deeper" by Jorge Cham, www.phdcomics.com, used by permission.

More generally, counterproductive stress is all too prevalent in our society. In education systems, it affects both instructors and students (Almeida et al., 2020; Tortella et al., 2021; Vogel & Schwabe, 2016). We can apply basic physics to help our students reduce unnecessary stress and better align actions with intentions. As one Nobel physics laureate famously said, "The first principle (of science) is that you must not fool yourself – and you are the easiest person to fool" (Feynman, 1974). Physics can serve as a powerful tool to help us perceive reality better, including within our own bodies (Park & Hawkins, 2023).

# Activities

The activities described below can engage students to experience the physics they are learning in their bodies in ways that can also help them reduce unnecessary stress. We invite fellow educators to play with and adapt these activities to a wide range of instructional contexts. In particular, each activity may be flexibly modified so that everyone can participate, including students and instructors with mobility differences.

## Newton's 3rd Law

Perhaps the simplest and most powerful example of embodying physics involves Newton's 3rd law, known colloquially as action-reaction. As fellow physics educators know, students new to physics usually have considerable difficulty with Newton's 3rd law, which we can overcome through guided learning activities (e.g. Sayre et al., 2012; Smith & Wittmann, 2007). We guide students to realize that a "force", as defined in physics, is a mutual interaction between two objects (Etkina et al., 2019). As Paul Hewitt put it, "you cannot touch without being touched" (Hewitt, 2006).

*In practice:* One simple activity adapted from the Online Active Learning Guide (Etkina et al., 2019) involves asking students to hold one arm out horizontally and angle their fingers upwards as far as they possibly can (see image (a) below). Then, invite students to press their hands against a wall, and ask them what they observe about how far their wrist can bend with the help of the wall. As in image (b) below, the wrist clearly bends further, almost at a right angle. We can then invite students to propose possible explanations for this observation.

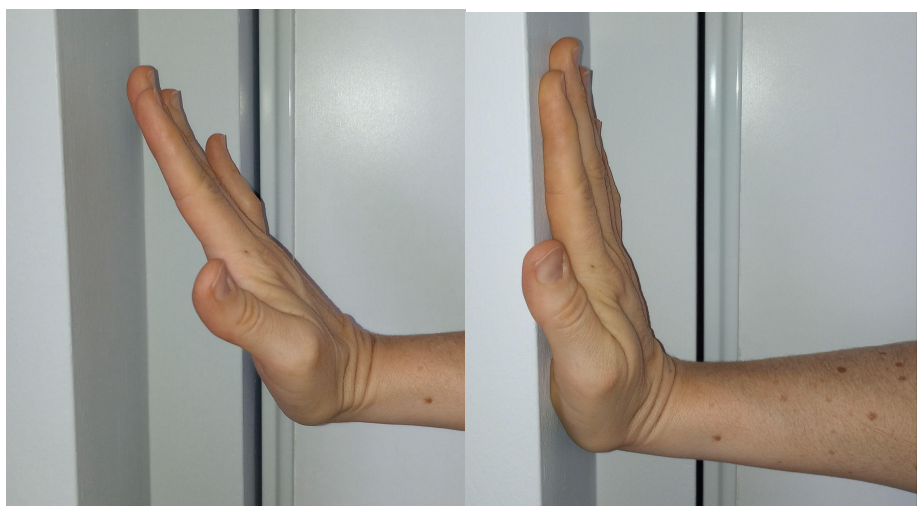

(a)  (b)

*Photos by the author.*

Most students initially tend to think of this action as "the person is pushing the wall," when really the wall and the person are both pushing together. In fact, the wall pushes back on the person's hand as much as the person's hand pushes on the wall. Both forces in this mutual interaction arise simultaneously. We can invite students to apply this to other simple actions and daily activities, such as standing or walking.

Often people think of standing as something we need to actively do by ourselves. Yet from Newton's 3rd law, we know intellectually that we cannot push down into the ground without the ground pushing up on us.
*In practice*: We can invite students to feel the force of the floor pushing up on them, just like they could feel the force of the wall pushing back on their hands. As they gently shift their weight back and forth (such as between their left foot and right foot), can they feel any shift in the simultaneous upwards support of the ground? We are never standing all by ourselves, but rather standing with the support of the ground.

This mutual interaction occurs within every part of our body: There is a mutual vertical interaction due to gravity between every differential bit of mass or "*dm*" and Earth, and there is a mutual electrostatic push or pull between adjacent bits. If we stand still, to keep us in equilibrium and supported, the electrostatic push or pull between adjacent body parts needs to exert just enough upwards force on each part to balance the downwards force exerted by Earth on that part. Note that we can never directly feel the force exerted by Earth on our body parts, but we can feel the mutual interaction of support between parts (as well the overall mutual support or "normal force" between our body and the ground).
*In practice:* We can invite students to feel this mutual interaction in their shoulders, for example. Leaning over at the waist, we can dangle our arms freely, so that it is easier to feel the force exerted by the arms pulling down on the shoulders and the shoulders supporting the arms. We can then gradually return to an upright standing position and try to feel that same dangling and support, that same mutual interaction at our shoulders. If we hold our arms more stiffly, we activate muscles that are pulling in opposite directions (e.g. up and down) to maintain equilibrium. Yet we can release that extra unnecessary muscle tension and still be in equilibrium, and then we can more easily feel the mutual interaction in our shoulders.

Most people tend to think of walking, too, as something we must do by ourselves. Especially when we're in a rush, we will activate extra muscles unnecessarily in an effort to walk faster. Yet we know that, to accelerate forward, we need to push backwards on the ground and it's the force of friction between the ground and our feet that pushes us forward.
*In practice:* We can invite students to feel that mutual interaction with the ground as they walk, to genuinely feel the ground pushing them forwards to start walking (and backwards to slow down). Fundamentally, we are working together with Earth and the ground in order to walk; we are never walking "alone".

Through such activities, students can start to feel the reality of mutual interaction over the illusion of one-way action, both overall between the body and the ground and also at a more refined level throughout the body. When students feel mutual interaction more, they can better appreciate Newton's 3rd Law, release unnecessary muscle tension, and more comfortably stand, walk, move, or dance.

## Pendula

Building on the arm-dangling activity above, we can invite students to feel the freedom of motion in swinging our arms, and how we don't need to keep pushing our arms through: we can let Earth pull them down and let them swing through on their own. We can also apply this to walking, feeling the freedom of swinging our legs and letting the ground catch us with each step.

## Centre of Mass

We can also invite students to experience the physics concept of the centre of mass in their bodies. As we stand and shift our centre of mass back and forth from left to right, we can feel the shift in the mutual support at each foot. Keeping the torso centred, as we lift an arm, or move both arms to one side, we can tune in to the subtle shift in the support at our feet. We can play with the position of our hips, and notice how our hip position shifts our centre of mass and makes it easier or harder to balance on one leg. Feeling such shifts in the body's centre of mass can help make moving more comfortable.

We can also invite students to apply the centre of mass to stand up more easily from sitting in a chair. If we sit back in the chair and keep our centre of mass over the chair, we cannot stand up; rather, we need to shift our centre of mass forward until it is over our feet in order to stand. Tuning into the mutual interaction with the ground as we stand is again helpful (bringing back Newton's 3rd Law). If students have older relatives who are starting to have difficulty standing up from sitting, they can use this to help them.

## Pressure

Lastly, if we are taking a deep breath or observing our breath, we can tune in to the air pressure around us. We don't have to force the air in to breathe in, rather we just make more space and the air will rush into our lungs.

These are but a few examples. We warmly invite you to play with how you and your students might experience physics in the body more, and use that experience to release unnecessary muscle tension while still doing the same everyday actions. If you would like, please share your explorations and questions via my email address above. Most of us have experienced that extra mental stress leads to extra physical muscle tension in our body. Yet the reverse also applies, and by releasing extra physical muscle tension (such as laying down the burden of standing or sitting to the ground and feeling the ground support us), we can experience a more relaxed and open mind – better able, for example, to learn even more fun physics!

# Acknowledgements

This article is based on a 2023 OAPT conference workshop facilitated by the author and co-designed with Mackenzie Hawkins of the Wuwei Tai Chi School, and integrates insights from retired physicist and tai chi master Dr. Wonchull Park (Park & Hawkins, 2023).
Thanks to Hawkins, Dr. Eugenia Etkina, Dr. Ann-Marie Pendrill, and Eric Haller for helpful comments to improve the article.